\documentclass[a4paper,aps,preprintnumbers,showpacs,twocolumn,superscriptaddress,nofootinbib,amsmath,amssymb]{revtex4}
\usepackage[dvips]{graphics}

\begin{document}
\title{An analytical approximation for the Einstein-dilaton-Gauss-Bonnet black hole metric}

\author{K.~D.~Kokkotas}
\email{kostas.kokkotas@uni-tuebingen.de}
\affiliation{Theoretical Astrophysics, Eberhard-Karls University of T\"ubingen, T\"ubingen 72076, Germany}
\author{R.~A.~Konoplya}
\email{roman.konoplya@uni-tuebingen.de}
\affiliation{Theoretical Astrophysics, Eberhard-Karls University of T\"ubingen, T\"ubingen 72076, Germany}
\affiliation{Institute  of  Physics  and  Research  Centre  of  Theoretical  Physics  and  Astrophysics, Faculty of Philosophy and Science, Silesian University in Opava, Opava, Czech Republic}
\author{A.~Zhidenko}
\email{olexandr.zhydenko@ufabc.edu.br}
\affiliation{Centro de Matem\'atica, Computa\c{c}\~ao e Cogni\c{c}\~ao,
  Universidade Federal do ABC (UFABC), Rua Aboli\c{c}\~ao, CEP:
  09210-180, Santo Andr\'e, SP, Brazil}

\begin{abstract}
We construct an analytical approximation for the numerical black hole metric of P. Kanti, et. al. [Phys.\ Rev.\ D {\bf 54}, 5049 (1996)] in the four-dimensional Einstein-dilaton-Gauss-Bonnet (EdGB) theory. The continued fraction expansion in terms of a compactified radial coordinate, used here, converges slowly when the dilaton coupling approaches its extremal values, but for a black hole far from the extremal state, the analytical formula has a maximal relative error of a fraction of one percent already within the third order of the continued fraction expansion. The suggested analytical representation of the numerical black hole metric is relatively compact and good approximation in the whole space outside the black hole event horizon.
Therefore, it can serve in the same way as an exact solution when analyzing particles' motion, perturbations, quasinormal modes, Hawking radiation, accreting disks and many other problems in the vicinity of a black hole. In addition, we construct the approximate analytical expression for the dilaton field.
\end{abstract}
\maketitle

\section{Introduction}

Observations of black holes through the gravitational waves they emit or by electromagnetic spectra of the surrounding matter are intensively developing in the past few years \cite{TheLIGOScientific:2016src,Goddi:2016jrs}.
Nevertheless, the current observational data, although compatible with the einsteinian gravity, leaves a large window for alternative theories \cite{Konoplya:2016pmh,Yunes:2016jcc}. The latter appear as attempts to solve a number of principal theoretical problem, such as construction of a consistent quantum gravity, the nature of singularities, the dark energy/dark matter problem, the hierarchy problem etc.

One of the most interesting and motivated alternative theories has the form of Einstein gravity with an added second order in curvature (Gauss-Bonnet) term, which is coupled to a scalar field, called dilaton. This theory, (thus, \emph{Einstein-dilaton-Gauss-Bonnet}) comes from the low-energy limit of string theory and, thereby, represents quantum corrections to Einstein gravity inspired by the string theory.
The black hole solution for this theory is unknown in an analytical form, however, the numerical solution was found by P. Kanti and co-workers \cite{Kanti:1995vq} for a static spherically symmetric case. Numerical solution for the rotating black hole has been recently obtained in \cite{Kleihaus:2016dui}, while perturbative solutions in terms of the rotation parameter were developed in \cite{Ayzenberg:2014aka,Maselli:2015tta}.

In addition to black holes, neutron star models in EdGB gravity have been also constructed both for the static and the slowly rotating case \cite{Panietal2011}. While their axial oscillations have been studied in \cite{Blzquez-Salcedo2016}. Rapidly rotating neutron star models were also constructed in EdGB gravity \cite{Kleihausetal2014,Kleihausetal2016}. Their I-Love-Q relations were derived and it was found that the deviations from pure general relativity are relatively small.

Recently, a number of potentially observable properties of the Einstein-dilaton-Gauss-Bonnet black holes have been considered. Reflection spectrum of accreting black holes was analyzed in \cite{Zhang:2017unx} and its quasi-period oscillations in \cite{Maselli:2014fca}. The shadows cast by the black hole were considered in \cite{Younsi:2016azx,Cunha:2016wzk}, while the gravitational quasinormal modes were calculated in \cite{Blazquez-Salcedo:2016enn,Pani:2009wy,Blazquez-Salcedo:2017txk}. At the same time, for a number of problems related to simulations of evolution of matter and radiation in the vicinity of a black hole, thermodynamics, Hawking radiation etc., the analytical expression for the metric, even if approximate, would be preferable.

An approach to finding such an analytical approximation is based on the generic parametrization for black hole-space times formulated in \cite{Rezzolla:2014mua} for spherically symmetric and in \cite{Konoplya:2016jvv} for axially symmetric black holes for arbitrary metric theories of gravity. For spherical symmetry this parametrization is based on the continued-fraction expansion of the metric coefficients in terms of a compactified radial coordinate \cite{Rezzolla:2014mua}, what provides the superior convergence of the expansion to an exact solution or accurate numerical data for the metric. The expansion is designed in such a way, that the coefficients in the continued fraction are fixed by behavior of a metric near the event horizon, while the pre-factors are introduced to match the asymptotic behavior at infinity. This way, the accurate analytical expression can be obtained for \emph{the whole space} outside the event horizon, and not only near the black hole or far from it, as it usually happened in various approaches which deform/perturb the Schwarzschild and Kerr spacetimes by a set of multipoles. In this way the analytical approximation for the numerical non-Schwarzschild asymptotically flat black hole solution was obtained for the Einstein-Weyl theory \cite{Kokkotas:2017zwt}. There, the expansion of the fourth order provided the maximal relative error of about fraction of one percent.

In the present work we shall construct an analytical approximation for the numerical solution obtained by P. Kanti and co-workers  \cite{Kanti:1995vq} for the four-dimensional spherically symmetric and asymptotically flat black hole in the Einstein-dilaton-Gauss-Bonnet theory of gravity. We shall use the continued fraction parametrization and show that in order to meet the maximal error of a few tenths of percent for the non-extremal black hole, it is sufficient to take first three orders of the expansion. For near extremal values of the dilaton coupling, which is still compatible with asymptotically flat metric and existence of the horizon, the continued fraction converges slowly, so that more than three orders must be taken to find a good approximation. The obtained metric has relatively compact form and can be effectively used for further simulations of accreting matter in its vicinity, analysis of Hawking radiation, quasinormal modes, etc.

The paper is organized as follows. Sec. II gives the essentials of the Einstein-dilaton-Gauss-Bonnet theory and the numerical solution obtained in  \cite{Kanti:1995vq}. Sec. III relates the continued fraction parametrization for this numerical solution and derivation of the analytical formula for the metric coefficients. Sec. IV discusses the accuracy of the obtained analytical metric through the direct comparison of metric coefficients obtained numerically and analytically and by consideration of characteristics of the orbital photon motion.

\section{Black holes in the Einstein-dilaton-Gauss-Bonnet theory}

The Lagrangian for dilaton gravity with a Gauss Bonnet term reads
\begin{eqnarray}
{\cal L}&=&\frac{1}{2}R - \frac{1}{4} \partial_\mu \phi \partial^\mu \phi \\\nonumber&&+ \frac{\alpha '}{8g^2} e^{\phi }\left(R_{\mu\nu\rho\sigma}R^{\mu\nu\rho\sigma} - 4 R_{\mu\nu}R^{\mu\nu} + R^2\right),
\end{eqnarray}
where $\alpha '$ is the Regge slope and $g$ is the gauge coupling constant.

Following \cite{Kanti:1995vq}, a shifting of the dilaton field function $\phi\rightarrow\phi-\ln(\alpha '/g^2)$ will lead to $\alpha '/g^2=1$. Furthermore, a spherically symmetric spacetime may be chosen
\begin{equation}
ds^2 = -e^{\Gamma(r)} dt^2 + e^{\Lambda(r)} dr^2 + r^2 (d\theta ^2 + sin^2 \theta d\varphi^2).
\label{metric}
\end{equation}
In this metric the functions $\phi(r)$, $\Gamma(r)$ and $\Lambda(r)$ defined as follows:
\begin{eqnarray}
&\phi''(r)&=-\frac{d_1(r, \Lambda, \Gamma', \phi, \phi')}{d(r, \Lambda, \Gamma', \phi, \phi')},
\label{thone}
\\
&\Gamma''(r)&=-\frac{d_2(r, \Lambda, \Gamma', \phi, \phi')}{d(r, \Lambda, \Gamma', \phi, \phi')},
\label{thtwo}
\\
&e^{\Lambda(r)}=&\frac{1}{2}\left(\sqrt{Q^2-6\phi'e^{\phi}\Gamma'}-Q\right),
\label{ththree}
\end{eqnarray}
where
$$Q(r)\equiv\frac{\phi'^2 r^2}{4}-1-\left(r+\frac{\phi' e^{\phi}}{2}\right)\Gamma' \, ,$$
while for the analytic description of the quite cumbersome expressions $d$, $d_1$, and $d_2$ the interested reader my check  the Appendix of \cite{Kanti:1995vq}.
It is noticeable that equation (\ref{thtwo}) is of first order for $\Gamma'(r)$ diverging at the event horizon $r_0$ as
$$\Gamma'(r)=\frac{1}{r-r_0}\left(1+{\cal O}(r-r_0)\right).$$
Therefore, a new function $\Psi(r)$ can be defined as
$$\Psi(r)\equiv\Gamma'(r)(r-r_0),$$
while the equations (\ref{thone}), (\ref{thtwo}) and (\ref{ththree}) can be solved numerically. Actually, the initial conditions will be specified  by requiring regularity of the functions $\Psi(r)$ and $\phi(r)$ at the event horizon
\begin{eqnarray}
&\phi(r_0)=&\phi_0,\nonumber\\
&\phi'(r_0)=&r_0 e^{-\phi_0}\left(\sqrt{1-6\frac{e^{2\phi_0}}{r_0^4}}-1\right) \, ,\label{initial}\\
&\Psi(r_0)=&1\, ,\nonumber
\end{eqnarray}
where $r_0$ and $\phi_0$ are arbitrary parameters. The above equations suggest  that the value of the dilaton on the event horizon $\phi_0$ must be always smaller than a particular value, in order to provide existence of the event horizon of a given radius:
\begin{equation}\label{limits}
e^{\phi_0} < \frac{r_0^2}{\sqrt{6}}\, .
\end{equation}

\begin{figure*}
\resizebox{\linewidth}{!}{\includegraphics*{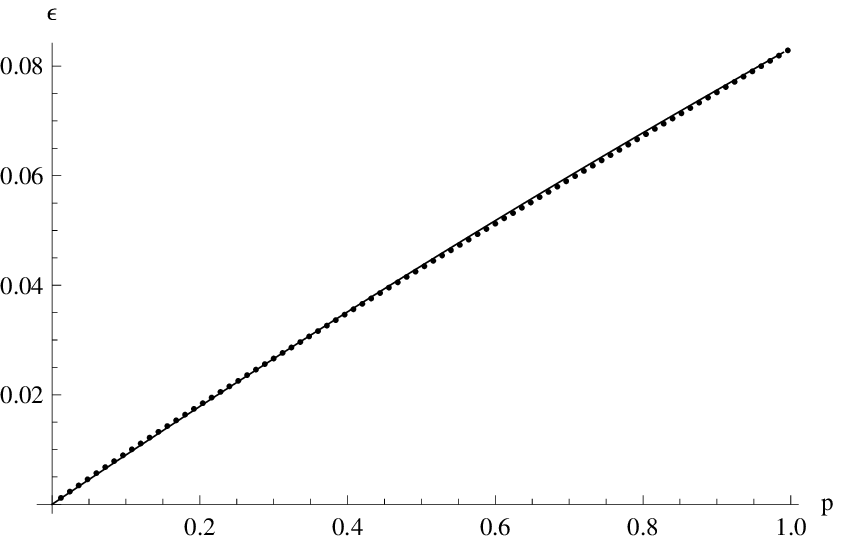}\includegraphics*{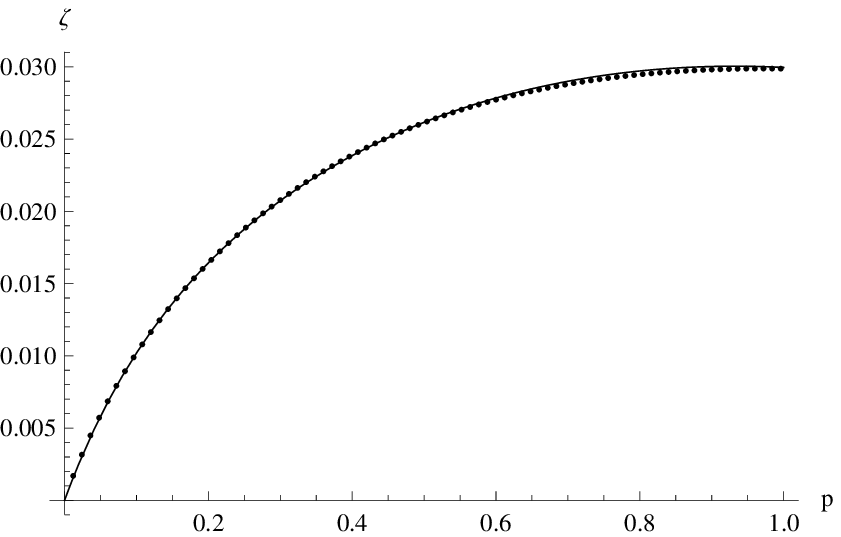}}
\caption{Fit (solid line) of the numerically found $\epsilon$  (left panel) and $\zeta$ (right panel) as functions of the parameter $p$. Notice that $\zeta(p)$ has a maximum at $p\approx0.97$.}\label{fig:epsilon-fit}
\end{figure*}

For each numerical solution (for a fixed set of all the parameters) the auxiliary function
$$S(r)=\intop_0^r\left(\frac{\Psi(r)-1}{r-r_0}+\frac{1}{r}\right)dr \, ,$$
is  also calculated numerically.  Actually, for large $r$ the function $S$ gets the asymptotic form
\begin{equation}
S(r)=S_{\infty}+\frac{r_0-2M}{r}+{\cal O}\left(\frac{1}{r^2}\right) \, .
\end{equation}
This allows one to read off the asymptotic mass $M$. Then,  the metric functions  $\Lambda(r)$ can be derived from equation (\ref{ththree}) while $\Gamma(r)$ will be extracted from the following relation
\begin{equation}
e^{\Gamma(r)}=\left(1-\frac{r_0}{r}\right)e^{S(r)-S_{\infty}} \, .
\label{gamma}
\end{equation}
In a similar manner one can read off the asymptotic dilaton parameters
\begin{equation}
\phi(r)=\phi_{\infty}+\frac{D}{r}+{\cal O}\left(\frac{1}{r^2}\right),
\end{equation}
where  $\phi_{\infty}$ is the asymptotic value of the dilaton and $D$ its charge. In order to restore the asymptotically vanishing dilaton, one should perform the inverse shifting through consideration of the function $\phi(r)-\phi_{\infty}$. Then, one gets
$$\frac{\alpha '}{g^2}=e^{\phi_{\infty}}.$$
This asymptotic parameter allows one to calculate also $\zeta$ defined in \cite{Ayzenberg:2014aka}
\begin{equation}\label{zetadef}
\zeta\equiv\frac{\alpha'^2}{16g^4M^4}=\frac{e^{2\phi_{\infty}}}{(2M)^4}\, .
\end{equation}

In order to simplify the analysis, we fix $r_0=1$ and measure the radial coordinate in the units of the event horizon radius, so that the family of the EdGB black hole solutions can be parametrized via the following dimensionless parameter
\begin{equation}\label{fampar}
p\equiv6e^{2\phi_0}=\frac{6\alpha'^2}{g^4r_0^4}e^{2(\phi_0-\phi_{\infty})}\, , \qquad 0\leq p<1 \,,
\end{equation}
so that $p=0$ corresponds to the Schwarzschild black hole. Owing to the above dilaton shifting, the shifted dilaton function goes to minus infinity  when $p \to 0$. This choice does not cause any problems for our purposes, because $e^{\phi(r)}$ remains finite.

It is more convenient to parametrize the family of the EdGB black holes using $p$ instead of $\zeta$, which is not a monotonous function of $p$ (see the right plot of Fig.~\ref{fig:epsilon-fit}). The latter leads to a branching of the solutions for large $\zeta$, i.e. to the existence of two black holes of different size, corresponding to the same parameters $M$ and $\zeta$ \cite{Torii:1996yi}. It has been recently shown in \cite{Blazquez-Salcedo:2017txk} that the additional branch, which in our notations corresponds to the values of $p\gtrsim0.97$, is linearly unstable.

Using equations (\ref{thone}), (\ref{thtwo}), and (\ref{ththree}), one can expand $e^{\phi(r)}$ and $e^{\Lambda(r)}$ near the horizon and calculate all the series coefficients in terms of the parameter $p$ defined in equation (\ref{fampar}). From (\ref{gamma}) we find the expansion for $e^{\Gamma(r)}$, which also depends on $M$ and $S_{\infty}$, calculated numerically for each value of $p$.

\begin{figure*}
\resizebox{\linewidth}{!}{\includegraphics*{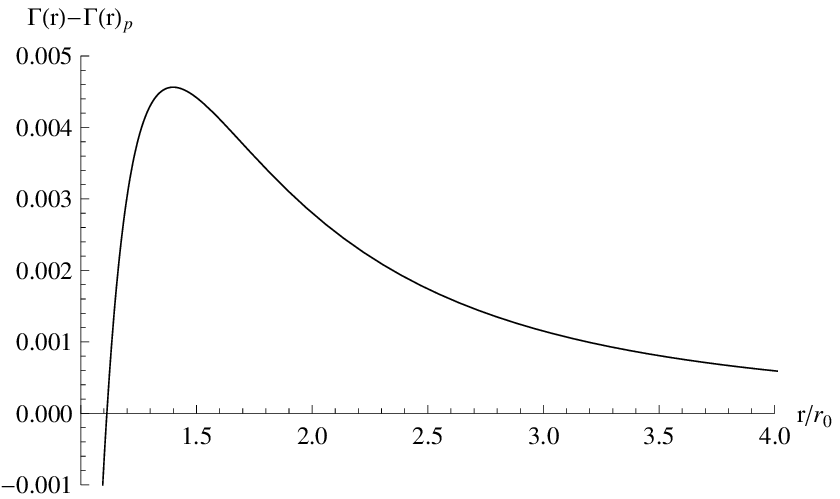}\includegraphics*{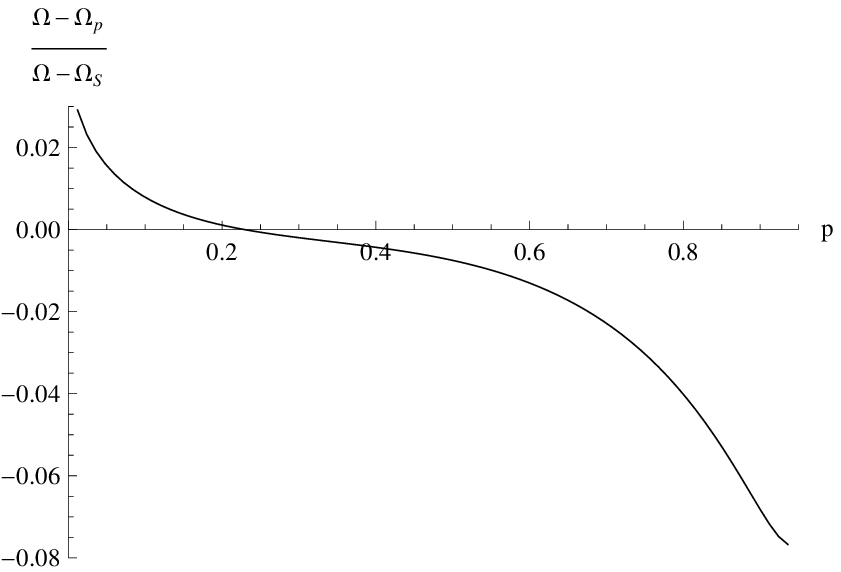}}
\caption{Left panel: difference between the numerical value of the function $\Gamma$ and its analytical approximation $\Gamma_P$ for $p=0.8$. The maximum deviation lies near the circular photon orbit. The relative error for the metric coefficient is $(e^{\Gamma(r)}-e^{\Gamma_p(r)})/e^{\Gamma(r)}\approx\Gamma(r)-\Gamma_p(r)$. Right panel: the difference between the numerical value of the value of the photon-orbit frequency $\Omega$ and the frequency calculated using the analytical approximation for the metric functions $\Omega_p$ divided by the difference between the frequency and its Schwarzschild value $\Omega_S=\frac{2}{3\sqrt{3}r_0}$. For $0<p<0.97$ the error is smaller by more then one order than the effect.}\label{fig:comparison}
\end{figure*}

\section{Analytical approximation}
Here it is advantageous to use the dimensionless compact coordinate $x$
\begin{equation}
x \equiv 1-\frac{r_0}{r}\,
\end{equation}
so that $x=0$ corresponds to the event horizon and $x=1$ to infinity.  Following the parametrization procedure given in \cite{Rezzolla:2014mua}, we define the new functions $A(x)$ and $B(x)$ through the following relations:
\begin{eqnarray}\label{Ax}
e^{\Gamma}&\equiv&xA(x)\,,\\
\label{Bx}
e^{\frac{\Gamma+\Lambda}{2}}&\equiv&B(x)\,.
\end{eqnarray}

We represent the above two functions as follows:
\begin{eqnarray}
A(x)&=&1-\epsilon (1-x)+(a_0-\epsilon)(1-x)^2+{\tilde A}(x)(1-x)^3,\nonumber\\
\label{asympfix}
B(x)&=&1+b_0(1-x)+{\tilde B}(x)(1-x)^2,
\end{eqnarray}
where ${\tilde A}(x)$ and ${\tilde B}(x)$ are given in terms of the continued fractions, in order to describe the metric near the event horizon $x=0$:
\begin{align}\nonumber
{\tilde A}(x)=\frac{a_1}{\displaystyle 1+\frac{\displaystyle
    a_2x}{\displaystyle 1+\frac{\displaystyle a_3x}{\displaystyle
      1+\frac{\displaystyle a_4x}{\displaystyle
      1+\ldots}}}}\,,\\\label{contfrac}
{\tilde B}(x)=\frac{b_1}{\displaystyle 1+\frac{\displaystyle
    b_2x}{\displaystyle 1+\frac{\displaystyle b_3x}{\displaystyle
      1+\frac{\displaystyle b_4x}{\displaystyle
      1+\ldots}}}}\,.
\end{align}

The coefficients $a_0$, $b_0$ and $\epsilon$ are introduced to match the post-Newtonian asymptotic at infinity. Thus,
\begin{eqnarray}
a_0 &=& \frac{(\beta -\gamma)(1+\epsilon)^2}{2},\\
b_0 &=& \frac{(\gamma-1)(1+\epsilon)^2}{2},
\end{eqnarray}
where $\beta$ and $\gamma$ are the corresponding PPN (Parameterized post-Newtonian) parameters \cite{Will:2005va}. The asymptotical behavior of the metric functions analysed in \cite{Kanti:1995vq} implies that $\beta=\gamma=1$, i.e. the post-Newtonian parameters for the static black hole solution coincide with those in General Relativity, leading to $a_0=b_0=0$. This is compatible with the current observational constrains. The asymptotic coefficient $\epsilon$ defines the relation between the position of the event horizon and asymptotic mass:
\begin{equation}
\epsilon=-\left(1-\frac{2M}{r_0}\right)\,.
\end{equation}
Here we certainly use the numerically found value of the mass $M$ in calculations.
Notice, that $\epsilon$ can be approximately expressed in terms of the dimensionless parameter $p$ defined in (\ref{fampar}). The accuracy is excellent as it can be seen in Fig.~\ref{fig:epsilon-fit} (left panel) while to leading order one can get
\begin{equation}\label{epsilonfit}
\epsilon\approx \frac{p}{11} - \frac{p^2}{131}\, .
\end{equation}

Once the radius of the event horizon is fixed, all quantities can be express as functions of the independent dimensionless parameter $p$.
Thus, the parameter $\zeta$ defined in equation (\ref{zetadef}) can be also  approximately expressed as a function of $p$ (see right panel of Fig.~\ref{fig:epsilon-fit})
\begin{equation}\label{zetafit}
\zeta\approx \frac{p}{1+4p}\left(\frac{3}{22}+\frac{2 p}{29}-\frac{p^2}{18}\right)\,,
\end{equation}
This approximate relation can be used to compare the black holes solutions with those in  \cite{Ayzenberg:2014aka} which are expressed in terms of $\zeta$.
For instance, we easily prove that
\begin{equation}
\epsilon\approx\frac{p}{11}+{\cal O}(p^2)=\frac{2}{3}\zeta+{\cal O}(\zeta^2),
\end{equation}
which is close to the accurate result \cite{Konoplya:2016jvv}
$$\epsilon=\frac{49}{80}\zeta+{\cal O}(\zeta^2).$$

Expanding (\ref{Ax}) and (\ref{Bx}) near the event horizon ($x=0$) we calculate numerically the coefficients $a_1, a_2, a_3,\ldots$, $b_1,b_2,b_3,\ldots$ for each value of $p$. Then, assembling data for each $p$ one can find an approximation for the coefficients $a_i$ and $b_i$ as functions of $p$ by fitting this numerical data. The optimal fitting will certainly depend on the order of our approximation, that is, on the order at which one truncates the continued fraction. At the third order of expansion, we find that the coefficients  $a_1$ and $b_1$ are best fit by the rational functions as follows
\begin{eqnarray}
a_1&=&\frac{5p}{(1-p)(5-3p)}\left(\frac{p^2}{40}+\frac{p}{19}-\frac{1}{13}\right)\,,\label{a1fit}\\
b_1&=&-\frac{13 p}{(1-p)(13-9p)}\left(\frac{p^2}{8}-\frac{5p}{13}+\frac{7}{27}\right)\,.\label{b1fit}
\end{eqnarray}
The coefficient $a_2$ crosses zero at $p\approx3/11$. We fit therefore $a_2$ as follows
\begin{equation}\label{a2fit}
a_2=\frac{3-11p}{(1-p)(2-p)}\left(\frac{15p}{19}-\frac{11}{13}\right).
\end{equation}

Unfortunately, in the parametric region where the absolute value of $a_2$ is small, the other coefficients are determined with significantly lower accuracy. That is why we find approximately
\begin{eqnarray}
b_2&=&-\frac{1}{(1-p)(5-4p)}\left(\frac{19 p^2}{12}+\frac{248 p}{19}-\frac{151}{10}\right)\,,\label{b2fit}\\
a_3&=&\frac{1}{1-p}\left(\frac{22}{9} - \frac{5 p}{7}\right)\,,\label{a3fit}
\end{eqnarray}
and truncate the higher coefficients, $b_3=0$, $a_4=0$. This approximates the metric functions for $p\lesssim0.97$ within a few tenths of percent (see Fig.~\ref{fig:comparison}). Unfortunately, the absolute values of the coefficients grow as $p\to1$ and for near-extreme values of $p$ the continued fractions (\ref{contfrac}) converge slower.

\begin{widetext}

\subsection{Analytical expressions for the metric functions}

The metric functions, obtained via the third order expansion described earlier, get the following form
\begin{equation} \label{final}
e^{\Gamma(r)}\approx \left(1-\frac{r_0}{r}\right)\frac{{\cal N}_1}{{\cal D}_1}
\quad \mbox{and} \quad
e^{({\Gamma(r)+\Lambda(r)})/{2}}\approx \frac{{\cal N}_2}{{\cal D}_2}\,,
\end{equation}
where
\begin{eqnarray}
{\cal N}_1&=& 30888 r r_0(r+r_0)(927r-1060r_0)p^6-3r_0(145693952r^3- 24067680r^2r_0 -156948260 r r_0^2-5338905r_0^3)p^5\nonumber\\
 &+& (3750946056r^4-3062334104r^3r_0-325162656r^2r_0^2 - 1478746401 r r_0^3 -53126788r_0^4)p^4 \nonumber \\
 &-& 2(6293682780r^4-7334803204r^3r_0-306613944r^2r_0^2 - 934415049 r r_0^3 +61245382r_0^4)p^3 \nonumber \\
 &+& 8(1350407212r^4-2160940683r^3r_0-64904931r^2r_0^2 - 139116640 r r_0^3 +62251200r_0^4))p^2\nonumber \\
 &+& 1048(1846581r^4 + 3798205r^3r_0 + 155610r^2r_0^2 + 270655 r r_0^3-321860r_0^4)p - 7666120r^3(509r-275r_0)\,, \nonumber \\
{\cal D}_1&=&11528 (1-p) (5-3 p) r^3 \left[ 117(927r-1060r_0)p^2-(74741r-121424r_0)p-67697r+36575r_0\right]\,,\nonumber
\\
{\cal N}_2&=&133380  r_0^2 p^4 - 7695  (58r^2+38 r r_0+75r_0^2)p^3 + 10 (471735 r^2-198819 r r_0 + 78964r_0^2) p^2 \nonumber \\
 &-& 4 (2398707r^2 - 1567647 r r_0 + 86450 r_0^2)p +26676 r (201 r-151r_0)\,,\nonumber \\
{\cal D}_2 &=&18 (13 -9p) r \left[95  (29r+19r_0) p^2- 60 (419 r-248 r_0)p + 6(3819 r-2869 r_0)\right]\,.\nonumber
\end{eqnarray}
\end{widetext}
\subsection{Analytical expression for the dilaton field}

In order to find an analytical approximation for the dilaton field we employ a similar approach, i.e. we define a new function
\begin{equation}\label{Fx}
e^{\phi-\phi_{\infty}}=F(x),
\end{equation}
which has the following form
\begin{equation}
F(x)=1+f_0(1-x)+{\tilde F}(x)(1-x)^2,
\end{equation}
where $f_0=D/r_0$ is the asymptotic coefficient and
\begin{equation}
{\tilde F}(x)=\frac{f_1}{\displaystyle 1+\frac{\displaystyle
    f_2x}{\displaystyle 1+\frac{\displaystyle f_3x}{\displaystyle
      1+\ldots}}}\,.
\end{equation}
Expanding (\ref{Fx}) near the event horizon ($x=0$) we calculate  numerically  the coefficients $f_1$, $f_2$, $f_3$, $\ldots$
The following fits can be found for the coefficients
\begin{eqnarray}
f_0&=&\frac{p}{1+35p}\left(\frac{48}{11}+\frac{222p}{19}-5p^2\right)\,,\label{f0fit}\\
f_1&=&\frac{p}{1+25p}\left(\frac{57}{17}+\frac{89p}{6}\right)\,,\label{f1fit}\\
f_2&=&\frac{1}{1-p}\left(\frac{8}{15}+\frac{13p}{8}-\frac{19p^2}{9}\right)\,,\label{f2fit}
\end{eqnarray}
while  the higher coefficients $f_3=0$ were truncated.

The previous analysis leads to the following analytical approximation for the dilaton

\begin{eqnarray}
e^{\phi(r)-\phi_{\infty}}&\approx& \frac{{\cal N}_3}{{\cal D}_3}\,,
\label{finalf}
\end{eqnarray}
where
\begin{widetext}
\begin{eqnarray}
{\cal N}_3 &=&-337535000  (r - r_0)r_0 p^6 + 88825(9853r-11653r_0)  r_0  p^5 \nonumber \\
 &+& 85(27797000 r^2 - 23776241 r r_0 + 9993021 r_0^2) p^4 -15(35832005r^2 - 67254703 r r_0+3991594r_0^2)p^3 \nonumber \\
& - & 68(25902415r^2 - 7037569 r r_0 + 1078359 r_0^2)p^2  -171(692835r^2 - 201467 r r_0 + 7672r_0^2)p \nonumber \\
&+& 85272  (23r+8r_0) r  \,, \nonumber\\
{\cal D}_3 &=&3553 (25 p+1)(35 p+1) r \left[760 ( r-r_0)p^2 - 45(5r-13r_0)p- 8(69 r-24 r_0)\right] \,. \nonumber
\end{eqnarray}
\end{widetext}

In the supplementary material we share with readers the Mathematica\textregistered{} notebook in which the obtained analytical formulas for the metric coefficients and the dilaton field are written down.

\section{Accuracy of the approximation: Circular photon orbits}

Now we would like to understand how good the found analytical metric (\ref{final}) approximates the accurate numerical solution of \cite{Kanti:1995vq}. The immediate comparison of the metric coefficients for numerical and analytical approaches show (see the left plot on Fig.~\ref{fig:comparison}) that the deviation from the numerical data is never higher than a few tenths of one percent. It is interesting that the maximal error occurs in the region where the innermost stable circular orbit, photon orbits, peak of the effective potential for quasinormal modes occur. In other words, the error is maximal in the most important region where all scattering processes occur. Therefore, we will pay special attention to this region and suggest another test of accuracy through comparison of orbit frequencies of photons calculated in the background of numerically derived metric and its analytical approximation (\ref{final}) derived in this work.

Thus, we shall study the circular photon orbits in the backgrounds of the numerical version of the metric and in its analytical approximation derived here. Owing to spherical symmetry, we can assume that the particle moves in the equatorial plane and take $\theta=\pi/2$. We shall associate $r_c$ with the radial coordinate of the particle in its geodesic motion along a circular orbit. We shall further estimate the maximal difference between the accurate and the approximate metrics. The coordinate $r_c$ satisfies the following equations
\begin{eqnarray}\nonumber
ds^2&=&-e^{\Gamma(r_c)}dt^2+r_c^2d\varphi^2=0,\\\nonumber
d^2r_c&=&\left(-\frac{1}{2} \Gamma'(r_c) e^{\Gamma(r_c)}dt^2+r_cd\varphi^2\right)e^{-\Lambda(r_c)}=0,
\end{eqnarray}
where the first equation is fulfilled for null geodesics, while the second one is for circular orbits.
The combination of the two equations leads to the following equation for $r_c$
\begin{equation}
r_c\Gamma'(r_c)=2,
\end{equation}
which can be solved numerically.

The corresponding orbital frequency $\Omega$, which is an observable quantity, is given by
\begin{equation}
\Omega=\frac{d\varphi}{dt}\Biggr|_{r=r_c}=\frac{1}{r_c}e^{\Gamma(r_c)/2}.
\end{equation}

Let us point out that the limiting value of the dilaton parameter (\ref{limits}) allows only a relatively small deviation of the numerical solution \cite{Kanti:1995vq}  from the Schwarzschild metric. Thus, it is important to ensure that the difference between the analytical approximation and the precise numerical data are much smaller than the difference between the Schwarzschild solution and the Einstein-dilaton-Gauss-Bonnet black hole metric of \cite{Kanti:1995vq}. In other words, it is essential that the effect of deviation from the Schwarzschild geometry owing to extra couplings are not ``absorbed'' by the error of our approximation owing to the truncation of the continued fraction. Therefore, we calculate the frequency for the analytically approximated metric and compare it with its precise value and the photon orbit frequency for the Schwarzschild black hole (see the right panel of Fig.~\ref{fig:comparison})
$$\Omega_S=\frac{2}{3\sqrt{3}r_0}.$$
\begin{figure*}
\resizebox{\linewidth}{!}{\includegraphics*{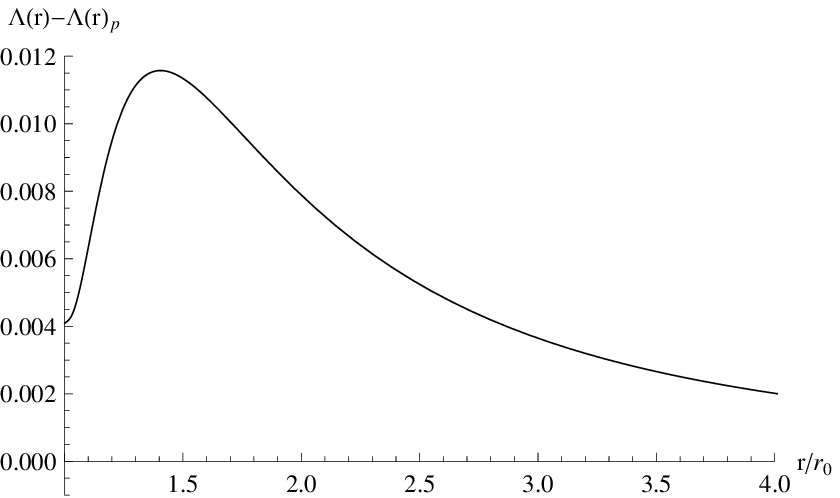}\includegraphics*{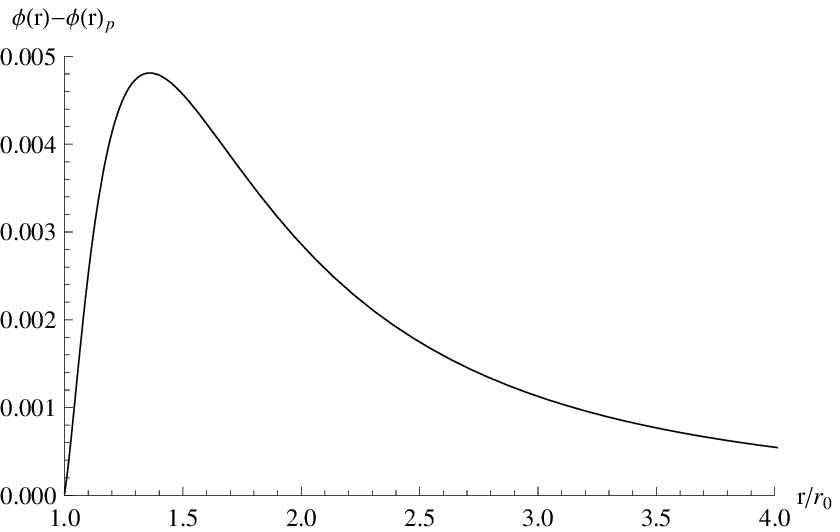}}
\caption{Difference between the numerical values of the functions $\Lambda$ (left panel) and $\phi$ (right panel) and their analytical approximations  for $p=0.5$. The maximum lies near the circular photon orbit. }\label{fig3:diff}
\end{figure*}
From the plots on Fig.~\ref{fig:comparison} one can see that the error due to the third order approximation leads to a deviation of \emph{at least} one order of magnitude smaller than the  deviation of the EdGB spacetime from the einsteinian geometry, when the black hole is far from its extremal state.
From Fig. \ref{fig3:diff} it can be seen that the maximal error  for the function $\Lambda(r)$ and the dilaton field $\phi(r)$ is of the same order as in  Fig.~\ref{fig:comparison}. Therefore, the found analytical metric (\ref{final}) can be effectively used for testing quantum corrections to the Einstein gravity.

\section{Final remarks}

In this article we present a method of construction an  analytical approximation to the metric of the  asymptotically flat and spherically symmetric four-dimensional Einstein-dilaton-Gauss-Bonnet black hole \cite{Kanti:1995vq}. An analytic approximate expression  for the dilaton field has been also derived.  Application of the continued fraction parametrization allowed us to find  a relatively compact form of the metric (\ref{final}), which, at the same time, provides good accuracy of a fraction of one percent. Therefore, the found analytical approximate metric can serve in the same way as an exact solution for analysis of various phenomena in the vicinity of a black hole, such as, particle motion, gravitational lensing, Hawking radiation, perturbations and quasinormal modes of black holes, scattering of fields, accretion of matter and others. In all the above phenomena it is essential to have a good approximation not only near the event horizon or/and infinity, but also in the intermediate region, on which all the scattering and accreting processes strongly depend. Here we met this requirement by providing a good analytical approximation in the whole space outside the event horizon.

An interesting test of our approximate analytical metric could be calculation of gravitational quasinormal modes and comparison of them with those found recently for the numerical metric  \cite{Blazquez-Salcedo:2017txk}. Although the quasinormal modes of a four-dimensional dilatonic black holes without Gauss-Bonnet term are well studied by now  (see, for example, \cite{Konoplya:2001ji,Fernando:2016ftj,Ferrari:2000ep,Kokkotas:2015uma}), the presence of  the Gauss-Bonnet term evidently leads to new phenomena and instability for some values of the parameters \cite{Blazquez-Salcedo:2017txk,Konoplya:2017lhs,Ahn:2014fwa,Zangeneh:2015tva}. The analytical approximation can also be obtained for  more general cases. For example, when the scalar field has an additional coupling \cite{Torii:1996yi}, which equals to unity in the case of the heterotic string theory we were limited here. Finally, our next aim in this direction is to find an analytical approximation for the numerical rotating EdGB black hole solution \cite{Kleihaus:2016dui}.

\acknowledgments{
R. K. was supported by ``Project for fostering collaboration in science, research and education'' funded by the Moravian-Silesian Region, Czech Republic and by the Research Centre for Theoretical Physics and Astrophysics, Faculty of Philosophy and Science of Sileasian University at Opava. A.~Z. thanks Conselho Nacional de Desenvolvimento Cient\'ifico e Tecnol\'ogico (CNPq) for support and Theoretical Astrophysics of Eberhard Karls University of T\"ubingen for hospitality.}

\end{document}